\documentclass[12pt]{article}
\usepackage{latexsym,amsfonts,amsmath,amsthm,amssymb}

\begin{document}

\title{Bell's inequality: Physics meets Probability}
\author{Andrei Khrennikov\\International Center for Mathematical Modeling \\in Physics,
Engineering and Cognitive Science, \\ V\"axj\"o University,
S-35195, Sweden\\ email: Andrei.Khrennikov@vxu.se}

\maketitle

\begin{abstract}
In this review we remind the viewpoint that violation of Bell's
inequality might be interpreted not only as an evidence of the
alternative -- either nonlocality or ``death of reality'' (under
the assumption the quantum mechanics is incomplete). Violation of
Bell's type inequalities is a well known sufficient condition of
probabilistic incompatibility of random variables -- impossibility
to realize them on a single probability space. Thus, in fact, we
should  take into account an additional interpretation of
violation of Bell's inequality -- a few pairs of random variables
(two dimensional vector variables) involved in the EPR-Bohm
experiment are incompatible. They could not be realized on a
single Kolmogorov probability space. Thus one can choose between:
a) completeness of quantum mechanics; b) nonlocality; c) `` death
of reality''; d) non-Kolmogorovness. In any event, violation of
Bell's inequality has a variety of possible interpretations.
Hence, it could not be used to obtain the definite conclusion on
the relation between quantum and classical models.
\end{abstract}


\section{Introduction}

This paper was stimulated by the recent review of  Genovese
\cite{GEN} devoted to EPR and EPR-Bohm experiments, Bell's
inequality, quantum nonlocality, realism and all those questions
which are nowadays intensively discussed, see
\cite{B}--\cite{Weihs} for foundations and e.g.
\cite{KHRP}--\cite{KHRP5}  for recent debates. Genovese presented
an interesting and deep analysis of these fundamental problems.
However, Genovese's presentation and conclusions were standard for
a typical physical presentation of the Bell's arguments.

 The aim of this paper is
to remind to the physical community (especially, its quantum
information part) that Bell's inequality is an important  point
where {\it probability theory meets physics.} Unfortunately, the
fundamental role of probability in this framework is missed. In
any event physicists try to escape coupling of mentioned
fundamental physical problems with foundations of probability
theory.

The situation in quantum physics  is such as it could be in
general relativity if the modern mathematical formalization of
geometry were ignored. For example, assume that one would not know
that geometry is not reduced to the Euclidean geometry (that there
exists e.g. the Lobachevsky geometry). In such a case by finding
``non-Euclidean behavior'' she might make the conclusion about
``death of reality''. In some way it would really be ``death of
reality'', but only Euclidean reality. Non-Euclidean local effects
might be also imagine as nonlocal Euclidean effects. In
probability theory the Kolmogorov probability model is an analogue
of the Euclidean geometry. In this review we present the viewpoint
that violation of Kolmogorovness might be interpreted as ``death
or reality'' or nonlocality. However, this is death of only
Kolmogorovian reality. Kolmogorov nonlocalty might be in fact
simply non-Kolmogorov locality.

We present results of  studies on the problem of {\it
probabilistic compatibility of a family of random variables.} They
were done during last hundred years. And they have the direct
relation to Bell's inequality. However, a priory studies on
probabilistic compatibility  have no direct relation to the well
known fundamental problems which are typically discussed by
physicists, namely, {\it realism and locality}
\cite{B}--\cite{Weihs}, see e.g. \cite{KHRP}--\cite{KHRP5}  for
recent debates. After a review on conditions of probabilistic
compatibility, we shall try to find traces of this probabilistic
research in physics, e,g,. parameters of measurement devices,
negative probabilities, detectors efficiency, fair sampling,
rejection of photon (Lamb's anti-photon, Santos' views).

We remark that our considerations would not imply that the
conventional interpretation of Bell's inequality
\cite{B}--\cite{Weihs} should be rejected. In principle, Bell's
conditions (nonlocality, ``death of reality'') could also be taken
into account. Our aim is to show that Bell's conditions are only
{\it sufficient, but not necessary} for violation of Bell's
inequality.

Therefore other interpretations of violation of this inequality
are also possible. Bell's alternative -- either quantum mechanics
or local realism -- can be extended -- either existence of a
single probability measure\footnote{By using the terminology of
modern probability theory one should speak about existence of a
single Kolmogorov probability space \cite{KOL}, \cite{G1}.} for
incompatible experimental contexts or quantum mechanics. We notice
that existence of such a single probability was never assumed in
classical (Kolmogorov) probability theory, but it was used by J.
Bell to derive his inequality (it was denoted by $\rho$ in Bell's
derivation). Therefore if one wants to use Bell's inequality, he
should find reasonable arguments supporting Bell's derivation,
roughly speaking:

\medskip

{\it Why do we use such an assumption in quantum physics, although
we have never used it in classical probability theory?}

\medskip

In classical probability theory researchers never try to put
statistical  data collected on the basis of different sampling
experiments into one single probability space. However, in quantum
mechanics we (at least Bell and his adherents)  try to do this.
From the point of view of ``the probabilistic opposition'' to the
conventional interpretation of violation of Bell's inequality, the
crucial problem of Bell's considerations was placing statistical
data collected in a few totally different experiments
(corresponding to different setting of polarization beam
splitters) in one probabilistic inequality. We remark that the
same  trick was done by Feynman \cite{FEY} in his probabilistic
analysis of the two slit experiment, However, Feynman remarked
that his analysis demonstrated violation of laws of classical
probability theory. \footnote{He had no idea about the rigorous
axiomatics of modern classical probability theory (the Kolmogorov
model). Therefore he wrote about violation of laws of classical
Laplacian probability.} And this is not surprising, because he
proceeded against all rules of the conventional probabilistic
practice. Kolmogorov emphasized from the very beginning \cite{KOL}
that any experimental context induces its own probability space.
But Feynman considered three different contexts: $C_{12}$ two
slits are open, $C_1$ only the first slit is open, $C_2$ only the
second slit is open. By some reason (I think simply because  of
lack of education in probability) he wanted to put such data
collected under three different contexts into a single probability
space. The impossibility to do this Feynman interpreted as
astonishing violation of laws of classical probability theory. To
solve this problem, he decided to assume (in accordance with
fathers of quantum mechanics, see e.g. Dirac \cite{DIR}) that each
quantum particle interferes with itself.\footnote{Such a
conclusion is in visible contradiction with his own path integral
approach to quantum mechanics. Of course, one might say (as
Feynman did) that these trajectories are virtual. However, in such
a case one should explain such a funny coincidence of prediction
of the quantum operator formalism with the result of integration
with respect to these totally nonphysical trajectories. Feynman
did not do this. It is clear that we do not know all historical
details. It might be that at the beginning Feynman assigned to
these trajectories more physical meaning. We only know that his
path integral approach was terribly criticized by Bohr.  Feynman
was able to sell his path integral approach to Bohr (and hence to
the quantum community) only through personal contacts with Pauli.
It is clear that the conflict with Bohr was not possible to
resolve in the framework of physical trajectories.} Bell did more
or less the same thing with the EPR-Bohm experiment. In contrast
to Feynman, Bell did not even see the probabilistic inconsistency
of his considerations. This story was presented in detail in my
book \cite{KHI} (which is definitely unreadable for physicists,
because of too much probability).

This paper is based on the results of research on the
probabilistic structure of Bell's inequality. We can call this
(very inhomogeneous) group of researchers ``probabilistic
opposition'' \footnote{Names are presented in the alphabetic
order. In spite the evident fact that physicists totally ignore
this probabilistic research, the ``probabilistic opposition'' is
permanently disturbed by conflicts on priority.}: Accardi
\cite{ACC}--\cite{ACC2}, Fine \cite{FINE},  Garola and Solombrino
\cite{GAR1}--\cite{GAR3},  Hess and Philipp \cite{HPL2},
Khrennikov \cite{KHI}, \cite{KH1}--\cite{KHF}, Kupczynski
\cite{Kup1}- \cite{Kup3}, Pitowsky \cite{PI}, \cite{PI1}, Rastal
\cite{RAST}, Sozzo \cite{SOZ}. On one hand, it is amazing that so
many people came to the same conclusion practically independently.
On the other hand, it is also amazing that this conclusion is not
so much known by physicists (even for mathematically interested
researchers working in quantum information theory). There is
definitely a problem of communication. I hope that this review
would inform physicists about some general mathematical ideas on
Bell's inequality.

I remark that (to my knowledge) only one physicist (and moreover
very good experimenter!), Klyshko \cite{Klyshko1}-
\cite{Klyshko1}, obtained similar conclusion -- independently from
mathematicians! He did not know anything about mentioned research
in probabilistic analysis of Bell's arguments. However, he also
pointed out to non-Kolmogorovness as a reasonable alternative to
nonlocality, ``death of reality'' and completeness of quantum
mechanics.

  I also point out
to a series of papers of Andreev and Manko et al.
\cite{Andreev1}-- \cite{Andreev5} who used quantum tomographic
\cite{Manko1}-- \cite{Manko5} interpretation Bell's inequality.
Although they have never emphasized ``non-Kolmogorovness'' of
their approach, it is evident that the quantum tomographic scheme
for Bell's inequality is based on a family of probability measures
associated with involved experimental contexts. It is impossible
to construct a single probability serving for the total collection
of statistical data.

Finally, we note that Volovich  \cite{Vol1}, \cite{Vol2}  pointed
out to the role of space variables in the EPR-Bohm studies (which
was surprisingly missed from modern considerations). He also
emphasized the crucial difference between the original EPR
experiment for correlations of positions and momenta and its Bohm
version for photon polarization or electron spin. Khrennikov and
Volovich demonstrated in the rigorous probabilistic framework that
the original EPR experiment for measurement of continuous
variables for entangled systems (position and momentum) differs
crucially from its EPR-Bohm version for measurement of discrete
variables for entangled systems (projections of polarization or
spin).  We discuss this point in the present review.

Since in this paper we shall discuss Bell's proof of its
inequality and its versions, we present (for reader's convenience)
these proofs in the appendix. We remark that the analysis of
probabilistic assumptions of Bell's arguments is extremely
important for modern quantum physics, especially quantum
information, cryptography and computing. Consequences of the
modern interpretation of violation of Bell's inequality for
foundations of quantum mechanics (and nanotechnology) are really
tremendous.  Hence, conditions of derivation of this inequality
should be carefully checked. In this paper considerations are
concentrated on the analysis of the possibility to use a single
probability distribution underlying two dimensional marginal
distributions predicted theoretically by quantum mechanics.

We note that the present experimental situation for the EPR-Bohm
experiment is very complicated. It is well known that Aspect et
al. \cite{AS}, see also  Weihs et al. \cite{Weihs0}, \cite{Weihs},
shown that Bell's inequality is really violated by the
experimental data collected in four experiments corresponding to
choices of different settings of polarization. However, recently
Adenier and Khrennikov found that it was not the end of this great
experimental story. By analyzing the data from the first
experiment that closed locality loophole, see  Weihs et al.
\cite{Weihs0}, we found that data contains anomalies of the
following type. The Bell's expression for correlations is a linear
combination of two dimensional probability distributions for
polarization. The astonishing fact is that these experimental two
dimensional probability distributions essentially differ from
those which are predicted by quantum mechanics theoretically, see
\cite{AKH}, \cite{AKH1}. In some mysterious way these anomalous
deviations are cancelled (by compensating each other) in the
Bell's expression for correlations. It is even more astonishing
that the same anomalies were present in statistical data from the
pioneer experiment of Aspect et al. \cite{AS} which differs
crucially by its technical realization from the experiment in
Weihs et al. \cite{Weihs0}. This fact about statistical anomalies
was not communicated in the article  \cite{AS}. However, it can be
found in the PhD-thesis of Alain Aspect, see \cite{AST}.

Therefore new cleaner experiments should be done in future. The
common opinion that Bell's arguments are totally confirmed by
experiments (with just such a purely technological problem as
inefficiency of detectors) is far from experimental reality.

Recently Bell-type inequalities for tests of  compatibility of
{\it nonlocal realistic models} with quantum mechanics were
derived, see Legget \cite{Legget}. They were generalized  and
tested experimentally by Gr\"oblacher et al. \cite{Grob}. It was
an important step toward unification of positions of ``orthodox
quantum community'' and mentioned above ``probabilistic
opposition.'' Physicists started to understand that Bell's
condition of locality played a subsidiary role in business  with
Bell-type inequalities. Even without the locality condition one
can obtain Bell-type inequalities. The crucial condition (as it
was pointed out by e.g. Fine \cite{FINE} and Rastal \cite{RAST},
see \cite{KHI} for detailed presentation) is the existence of a
single probability measure serving all experimental settings
involved in a Bell-type inequality -- probabilistic compatibility
of random variables. Unfortunately, the latter point of view was
again missed by Legget \cite{Legget} and Gr\"oblacher et al.
\cite{Grob}, see section 14.

\section{Studies of Boole and Vorobjev on probabilistic compatibility}

 Consider a system of three random variables $a_i, i=1,2,3.$ Suppose for simplicity that they take discrete values and moreover
they are dichotomous: $a_i= \pm 1.$
Suppose that these variables as well as their pairs can be measured and hence joint probabilities for pairs are well defined:
$$
P_{a_i, a_j}(\alpha_i, \alpha_j) \geq 0
$$
and
$$
\sum_{\alpha_i, \alpha_j=\pm 1}
P_{a_i, a_j}(\alpha_i, \alpha_j)=1.
$$

\medskip

{\bf Question:} {\it Is it possible to construct the joint probability distribution,
$P_{a_1, a_2, a_3}(\alpha_1, \alpha_2, \alpha_3),$ for any triple of random variables?}

\medskip

Surprisingly this question was asked and answered for hundred years ago by Boole
(who invented Boolean algebras). This was found by Itamar Pitowsky \cite{IPT1}, \cite{IPT2},
see also preface \cite{KHRP2}. To study this problem,
Boole derived inequality which coincides with the well known in physics Bell's inequality.
Violation of this Boole-Bell inequality implies that for such a system of three random variables
the joint probability distribution  $P_{a_1, a_2, a_3}(\alpha_1, \alpha_2, \alpha_3)$ does
not exist.

\medskip

{\bf Example.} (see \cite{VR}) Suppose that
$$
P(a_1= +1, a_2=+1) = P(a_1=-1, a_2=-1) =1/2;
$$
$$
P(a_1= +1, a_3=+1) = P(a_1=-1, a_3=-1) =1/2;
$$
$$
P(a_2= +1, a_3=-1) = P(a_2=-1, a_3=+1) =1/2.
$$
Hence, $P(a_1= +1, a_2=-1) = P(a_1=-1, a_2=+1) =0;\; P(a_1= +1,
a_3= - 1) = P(a_1=-1, a_3= +1)=0, \; P(a_2= +1, a_3= +1) =
P(a_2=-1, a_3=-1) =0.$ Then it is impossible to construct a
probability measure which would produce these marginal
distributions.  We can show this directly \cite{VR}. Suppose that
one can find  a family of real constants $P(\epsilon_1,\epsilon_2,
\epsilon_3), \epsilon_j=\pm 1,$ such that
$$
P(\epsilon_1,\epsilon_2, +1) + P(\epsilon_1,\epsilon_2, -1)=
P(a_1=\epsilon_1 , a_2=\epsilon_2),...,
$$
$$
P(+1,\epsilon_2,\epsilon_3) + P(-1,\epsilon_2,\epsilon_3)=
P(a_2=\epsilon_2 , a_3=\epsilon_3).
$$
Then he immediately finds that some of these numbers should be
negative. In a more fashionable way one can apply Bell's
inequality for correlations, see appendix:$ \vert \langle a_1,
a_2\rangle - \langle a_2, a_3\rangle \vert \leq 1- \langle a_1,
a_3\rangle.$  We have:
$$
\langle a_1, a_2\rangle=1; \; \langle a_1, a_3\rangle=1; \langle
a_2, a_3\rangle= -1.
$$
Bell's inequality should imply: $1- (-1)=2 \leq 1-1=0.$ We remark
that in accordance with Boole we consider Bell's inequality just
as a necessary condition for probabilistic compatibility.

Thus Bell's inequality was known in probability theory. It was
derived as a constraint  which violation implies nonexistence of
the joint probability distribution.

Different generalizations of this problem were studied in probability theory. The final solution
(for a system of $n$ random variable) was obtained by Soviet mathematician Vorobjev \cite{VR} (as was found
by Hess and Philipp  \cite{HPL2}).  His result was applied in purely macroscopic situations --
in game theory and optimization theory.

We emphasize that for mathematicians consideration of Bell's type inequalities did not induce
revolutionary reconsideration of laws of nature. The joint probability distribution does not exist
just because those observables could not be measured simultaneously.

\section{The EPR-Bohm Experiment, Impossibility to Measure Three Polarization Projections Simultaneously}

We consider now one special application of Boole's theorem the EPR-Bohm experiment for measurements
of spin projections for pairs of entangled photons.\footnote{Although both Boole's and Bell's theorems
are based on the same inequality, the conclusions are totally different. These are
``nonexistence of the joint probability distribution'' and ``either local realism or quantum mechanics'',
respectively. Thus we would like to analyze the EPR-Bohm experiment from the viewpoint of Boole (Vorobjev,
Accardi, Fine, Pitowsky, Rastal, Hess and Philipp and
the author).} Denote corresponding random variables by
$a_\theta^{1}$ and $a_\theta^{2},$ respectively (the upper index $k=1,2$ denotes
observables for corresponding particles in a pair of entangled photons).
Here $\theta$ is the angle parameter determining
the setting of polarization beam splitter. For our purpose it is sufficient to consider three different
angles: $\theta_1, \theta_2, \theta_3.$ (In fact, for real experimental tests we should consider four
angles, but it does not change anything in our considerations).

By using the condition of precise correlation for the singlet state
we can identify observables $$a_\theta(\lambda) \equiv a_\theta^{1}(\lambda)=a_\theta^{2}(\lambda).$$
The following discrete probability distributions are well defined: $P_{a_\theta}(\alpha)$ and
$P_{a_{\theta_i}, a_{\theta_j}}(\alpha, \beta).$ Here $\alpha, \beta= \pm 1.$
We remark that in standard derivations of Bell's type inequality for probabilities (and not correlations), see
appendix, there are typically used the following symbolic expressions of probabilities:
$P (a_{\theta}(\lambda)= \alpha)$ and
$P(a_{\theta_i}(\lambda)= \alpha,  a_{\theta_j}(\lambda)=\beta).$ However, by starting with a single probability
$P$ (defined on a single space of ``hidden variables'' $\Lambda)$ we repeat Bell's schema (which we would not like to repeat in this paper).

Thus we are precisely
in the situation which was considered in probability theory. Boole (and Vorobjev) would ask: Do
polarization-projections for any triple of angles have the joint probability distribution? Can one
use a single probability measure $P?$
The answer is negative  -- because the Boole-Bell inequality is violated (or because necessary condition of Vorobjev
theorem is violated). Thus it is impossible to introduce the joint probability distribution for
an arbitrary triple of angles.

On the other hand, Bell started his considerations with the assumption that such a single  probability measure
exists, see appendix. He represented all correlations as integrals with respect to the same probability measure
$\rho:$
$$
\langle  a_{\theta_i}, a_{\theta_j} \rangle= \int_\Lambda a_{\theta_i}(\lambda)
a_{\theta_j}(\lambda) d P(\lambda).
$$
(We shall use the symbol $P,$ instead of  Bell's $\rho$ to denote probability).

In opposite to Bell, Boole would not be so much excited by evidence of violation of Bell's
inequality in the EPR-Bohm experiment. The situation when pairwise probability distributions exist.
but a single probability measure $P$ could not be constructed is rather standard.
What would be a reason for existence of $P$ in the case when the simultaneous measurement
of three projections of polarization is impossible?

A priory nonexistence of $P$ has nothing to do with nonlocality or ``death or reality.''
The main problem is not the assumption that polarization projections are represented
in the  ``local form'':  $$a_{\theta_i}^{1}(\lambda), a_{\theta_j}^{2}(\lambda)$$
and not in the ``nonlocal form'' $$a_{\theta_i}^{1} (\lambda \vert a_{\theta_j}^{2}=\beta),
a_{\theta_j}^{2}(\lambda \vert a_{\theta_i}^{1} =\alpha),$$ where $\alpha, \beta= \pm 1.$
The problem is nor assigning
to each $\lambda$ the definite value of the random variable -- ``realism.''

The problem is impossibility to realize three random variables
$$a_{\theta_1}(\lambda), a_{\theta_2}(\lambda), a_{\theta_3}(\lambda)$$ on the same
space of parameters $\Lambda$ with same probability measure $P.$ By using the modern
terminology we say that it is impossible to construct a Kolmogorov probability space
for such three random variables.

In this situation it would be reasonable to find sources of nonexistence of a Kolmogorov
probability space. We remark that up to now we work in purely classical framework-- neither the
$\psi$-function or noncommutative operators were considered. We have just seen \cite{AS}, \cite{Weihs0},
\cite{Weihs} that experimental
statistical data violates the necessary condition for the existence of a single probability $P.$
Therefore it would be useful to try to proceed purely classically in the probabilistic analysis of the
EPR-Bohm experiment. We shall do this in the next section.

\section{Contextual Analysis of the EPR-Bohm Experiment}

As was already emphasized in my book \cite{KHI}, the crucial point
is that in this experiment one combine statistical data collected
on the basis of  three different complexes of physical conditions
(contexts). We consider context $C_1$ -- setting $\theta_1,
\theta_2,$ context $C_2$ -- setting $\theta_1, \theta_3,$ and
finally  context $C_3$ -- setting $\theta_2, \theta_3.$ We recall
that already in Kolmogorov's book \cite{KOL} (where the modern
axiomatics of probability theory was presented) it was pointed out
that each experimental context determines its own probability
space. By Kolmogorov in general three contexts $C_j, j=1,2,3,$
should generate three Kolmogorov spaces: with sets of parameters
$\Omega_j$ and probabilities $P_j.$

The most natural way to see the source of appearance of such spaces is to pay attention to the fact
that (as it was underlined by Bohr) the result of measurement is determined not only by the initial state
of a system (before measurement), but also by the {\it whole measurement arrangement.} Thus states of measurement devices
are definitely involved. We should introduce not only space $\Lambda$
of states of a system (a pair of photons), but also spaces of states of polarization beam splitters --
 $\Lambda_\theta.$
(We proceed under the assumption that the state of polarization beam splitter depends only on the orientation
$\theta.$ In principle, we should consider two spaces for each $\theta$ for the first and the second splitters.
In reality they are not identical.) Thus, see \cite{KHI}, for the context $C_1$ the space of parameters (``hidden variables'')
is given by $$\Lambda_1= \Lambda \times \Lambda_{\theta_1}\times \Lambda_{\theta_2},$$
for the context $C_2$ it is $$\Lambda_2= \Lambda \times \Lambda_{\theta_1}\times \Lambda_{\theta_3},$$
for the context $C_3$  it is
$$\Lambda_3= \Lambda \times \Lambda_{\theta_2}\times \Lambda_{\theta_3}.$$ And, of course,
we should consider three probability measures $$dP_1(\lambda,\lambda_{\theta_1}, \lambda_{\theta_2}),
dP_2(\lambda,\lambda_{\theta_1}, \lambda_{\theta_3}), dP_3(\lambda,\lambda_{\theta_2}, \lambda_{\theta_3}).$$
Random variables are functions on corresponding spaces
$$a_{\theta_1}(\lambda,\lambda_{\theta_1}),
a_{\theta_2}(\lambda,\lambda_{\theta_2}),
a_{\theta_3}(\lambda, \lambda_{\theta_3}).$$ Of course, Bell's ``condition of locality'' is satisfied
(otherwise we would have e.g. $a_{\theta_1}(\lambda,\lambda_{\theta_1}, \lambda_{\theta_2}),
a_{\theta_2}(\lambda,\lambda_{\theta_2}, \lambda_{\theta_1})$ for the context $C_1).$

In this situation one should have strong arguments to assume that these three probability
distributions could be obtained from a single probability measure $$dP_1(\lambda,\lambda_{\theta_1}, \lambda_{\theta_2},
\lambda_{\theta_3})$$ on the space $$\Lambda= \Lambda \times \Lambda_{\theta_1}\times
\times \Lambda_{\theta_2} \times \Lambda_{\theta_3}.$$

\section{Consequences for quantum mechanics}

Finally, we come to quantum mechanics. Our contextual analysis of the EPR-Bohm
experiment implies that the most natural explanation of
nonexistence of a single probability space is that {\it the wave function does not determine
probability in quantum mechanics} (in contrast to Bell's assumption). We recall that Born's rule
contains not only the $\psi$-function but also spectral families of commutative operators which
are measured simultaneously. Hence, the probability distribution is determined by the $\psi$-function
as well as spectral families, i.e., observables.

Such an interpretation of mathematical symbols of the quantum formalism
does not imply neither nonlocality nor ``death of reality.''\footnote{One should not accuse the
author in critique  of J. Bell. J. Bell by himself did a similar thing with the von Neumann no-go theorem [], see
[], by pointing out that some assumptions of von Neumann were unphysical.}

\section{Bell's Inequality and Negative and P-adic Probabilities}

By looking for a trace in physics of the Boole-Vorobjev conclusion
on nonexistence of probability one can find that this problem was
intensively discussed, but in rather unusual form (at least from
the mathematical viewpoint).  During our conversations on the
probabilistic structure of Bell's inequality Alain Aspect
permanently pointed out to a probabilistic possibility to escape
Bell's alternative: either local realism or quantum mechanics.
This possibility mentioned by Alain Aaspect is consideration of
negative valued probabilities. A complete review on solving
``Bell's paradox'' with the aid of negative probabilities was done
by  Muckenheim \cite{Muc}. Although negative probabilities are
meaningless from the mathematical viewpoint (however, see
\cite{KHP}- \cite{KHP7} for an attempt to define them
mathematically by using $p$-adic analysis), there is some point in
consideration of negative probabilities by physicists. In the
light of our previous studies this activity can be interpreted as
a sign of understanding that ``normal probability distribution''
does not exist. Surprisingly, but negative probability approach to
Bell's inequality can be considered as a link to Boole-Vorobjev's
viewpoint on violation of Bell's inequality. Of course, the formal
mathematical description by using negative probabilities does not
have any reasonable physical interpretation

\section{Detectors Efficiency}

Another trace  of nonexistence of a single probability space can
be found in physical literature on detectors efficiency
\cite{PEA}-- \cite{Gill}. Theoreticians as well as experimenters
are well aware about the fact that the real experiments induce
huge losses of photons. Even if one associates (as Bell did) one
fixed probability distribution with the source (the initial
state), there are no reasons to assume that it is preserved by
detectors.\footnote{We recall that our basic hypothesis is that
Bell's inequality is violated due to non-Kolmogorovness.  We
discuss different sources of nonexistence of a single probability
space. One of such possible sources is inefficiency of detectors.}

\medskip

How reasonable is this attempt to explain violation of Bell's
inequality by inefficiency of detectors?

\medskip

I discussed this problem with many outstanding physicists. The
common viewpoint was expressed in the reply of Alain Aspect to my
question. He did not believe that interference-like behavior of
correlations in the two slit experiment is just a consequence of
inefficiency of detectors. For him (as well as for majority of
physicists) violation of Bell's inequality is a consequence of
fundamental quantumness of the EPR-Bohm experiment and not at all
a technological problem of detectors efficiency. Similar viewpoint
was presented by Philip Pearle,  the first who paid attention to
the possibility to simulate the EPR-Bohm correlations via
detectors inefficiency \cite{PEA}.

Now we would like to explore another viewpoint to the EPR-Bohm
correlations.  If one accept that the  correlations given by the
EPR-Bohm experiment are nothing else than a special exhibition of
the general interference phenomenon, then it would be surprising
that for photons interference is generated by inefficiency of
detectors, but for e.g. electrons it is a consequence of another
hidden mechanism (we proceed our discussion under the assumption
that quantum mechanics is not complete).\footnote{We remark that
rather common viewpoint is that the EPR-Bohm experiment in the
Bell's framework  is essentially new experiment comparing with the
two slit interference experiment. It is often mentioned ``old
quantum mechanics'' before Bell and ``new quantum mechanics''
coupled to violation of Bell's inequality. However, purely
mathematically these correlations can be obtained as a special
case of interference of probabilities, see \cite{KHINTER}. The
point of view that fundamentally the EPR-Bohm experiment in the
Bell's framework and the two slit experiment demonstrate the same
physical phenomenon -- interference -- was presented by a number
of theoreticians and experimenters \cite{KHRP}-- \cite{KHRP5}.}

Therefore, although  improvement of detectors efficiency is very
important for quantum foundations \cite{ED1}, \cite{ED2} one could
not expect that the EPR-Bohm-Bell paradox would be resolved via
approaching approximately  $100\%$ detectors efficiency.

\section{Fair sampling}

The assumption of fair sampling is typically misidentified with
detectors efficiency. However, they are essentially different.
Unfair sampling could take place even for detectors having $100\%$
efficiency.

By the fair sampling assumption ensembles of pairs
$\omega=(\omega_1, \omega_2)$ of output photons from two
polarization beam splitters have the same probabilistic properties
independently on orientations of splitters. Thus one can operate
with a single probability distribution.

However, a priory there are no reasons to assume that ensembles of
pairs of photons which pass polarization beam splitters for
different choices of orientations (and were identified as pairs by
using time window, see e.g. \cite{} for details) have identical
statistical properties, see e.g. \cite{AKH},\cite{AKH1}. Unfair
sampling implies that  a single probability which would serve all
orientations does not exist.

{\it Of course, one should construct a physical model of unfair
sampling process which might be performed by polarization beam
splitters.} Moreover, one should explain, cf. with remark on
detectors efficiency, why the photon interference is a consequence
of unfair sampling, but e.g. electron interference is a
consequence of something else.

Finally, we remark that  in general nonexistence of probability is
not reduced to inefficiency of detectors or unfair sampling.

\section{Extended semantic realism}

This is a generalization of quantum formalism, see Garola and
Solombrino \cite{GAR1}--\cite{GAR3},  Sozzo \cite{SOZ},  by which
each quantum observable gets an additional point to its spectrum,
say $a_0,$ denoting the event of nonregistration. In principle,
one could consider extended semantic realism as a possible
formalization of unfair sampling. However, this approach suffers
from the same problem as the efficiency detectors viewpoint. One
should find a physical model explaining nonobservations of non
negligible subensembles of systems.

\section{Anti-photon interpretation of violation of Bell's inequality}

All real experiments demonstrating violation of Bell's inequality
and at the same time guaranteing locality have been done for
photons. However, since first days of quantum mechanics,  many
prominent quantum physicists criticized Einstein's proposal of
photon, e.g. Lande \cite{L1}, \cite{L2} and Lamb \cite{L3}. For
example, Lamb wrote in his ``Anti-photon'' \cite{L3} p. 221: ``It
is high time to give up the use of the word ``photon'', and of a
bad concept which will shortly be a century old. Radiation does
not consists of particles ...''

The idea about purely wave structure of classical as well as
quantum light has interesting consequences in the EPR-Bohm
experimental framework. If one rejects the conventional picture of
detectors registering  particles (photons) and if one considers
detectors as devices integrating continuously (up to a certain
threshold) electromagnetic radiation, then the whole Bell's
representation loses its meaning. We could not associate hidden
variables to clicks of detectors. We could not consider an
ensemble of photons produced by a source and the corresponding
probability distribution.  This  idea was explored in different
versions  by Santos \cite{S1},\cite{S2} , Thompson,  Kracklauer
\cite{KR1}, \cite{KR2} and Roychoudhuri.

It is typically used the semiclassical model: light is classical,
but atoms are quantum. In this model the electromagnetic field is
not quantized by itself, but exchange  of energy is performed by
discrete portions -- quanta. It is not easy to reject completely
the idea that violation of Bell's inequality implies simply that
one should use the semiclassical model, instead or the completely
quantum one. All detectors are based on either scattering of
electrons by photons (photomultipliers tubes -- PMTs) from a
photodiode or creation by photons pairs electron-hole (avalanche
photodiodes -- APDs and the visible light counters -- VLPCs), see
\cite{ED1} for a detailed review. Thus photon-like discreteness of
counting might be just an illusion induced by  discreteness of
electron emission. The latter one might be explained by the
semiclassical model (as well as e.g. the photoelectric effect).

If detectors interact with the continuous electromagnetic field
then the picture statistical ensembles of photons which was used
by Bell is misleading. For example, the signal field could produce
photon-like counts via combination with noise and even vacuum
fluctuations.

New possibilities to test the anti-photon interpretation of
violation of Bell's inequality is provided  by Tungsten-based
Superconducting Transition-Edge Sensors (W-TESs), see \cite{ED2}.
These are ultra-sensitive microcalorimeters. It seems that such
detectors absorb even portions of photons (if the latter exist).
In contrast to PMTs, APDs and VLPCs, W-TESs functioning is not
based on the  threshold principle. It seems that W-TESs provide
access directly to energy of prequantum classical electromagnetic
field (of course, if such nonquantized field really exists). PMTs,
APDs and VLPCs react only to integral portions of energy $E_n=n
h\nu,$ where $n$ is the number of photons in the pulse and $\nu$
is the frequency of light (in accordance with quantum theory
$E=h\nu$ is photon energy). In the anti-photon framework a  pulse
can contain some random portions of photon, $E_n(\omega)=E_n +
\xi(\omega),$  where $\xi(\omega)$ is a random variable (here
$\omega$ is a random parameter) and $\vert \xi(\omega) \vert  < h
\nu.$ In principle, the EPR-Bohm correlations can be reproduced as
the result of cutoff of this random term. We emphasize that
$\xi(\omega)$ is not detector's noise. This is a part of the
original signal. Of course, $\xi(\omega)$ can interact with
noises. If detector's noise is not so high, then one can hope to
extract ``nonquantum part'' of the signal. If a special
(``noclassical character'') of the EPR-Bohm correlations was
really a consequence of detection cutoff, then by taking into
account this term we might expect to reproduce classical
correlations which would not violate Bell's inequality.

\medskip

The main problem of the anti-photon interpretation of violation of
Bell's inequality is impossibility to generalize this argument to
massive particles, cf. inefficiency of detectors and unfair
sampling. Adherents  of this interpretation, e.g. Santos
\cite{S1}, \cite{S2}, typically point out that local experiments
with massive particles violating Bell's inequality have never been
done.  This is an important argument. The majority of
experimenters see the locality loophole in the famous Boulder
experiment \cite{BOULD}.

However, if one accepts the viewpoint that the EPR-Bohm experiment
is a special case of interference experiments, then he should also
explain why the Copenhagen postulate on wave-particle duality
could not be applied to light, but it should be applied to massive
particles -- to explain interference of such particles.

We remark that Alfred Lande \cite{L1}, \cite{L2} presented
detailed description of the interference effects for massive
particles without using the wave-particle duality. For him massive
particles are just particles, but electromagnetic field is just
field. If one generalizes Lande's argument to the EPR-Bohm
experiment, he should be able to obtain the EPR-Bohm correlations
for the electron spin by using the purely particle picture.

Thus the only possibility to interpret the EPR-Bohm experiment by
rejecting Bohr's principle of complementarity (and hence the
Copenhagen interpretation) is to create a purely wave model of the
EPR-Bohm correlations for experiments with light and a purely
particle model for experiments with massive particles.

\section{Anomalies in experimental data}

 We found \cite{AKH}, \cite{AKH1} that the experimental correlations for photon
polarization have an intriguing property. In the experimental data
there are visible non-negligible deviations of ``experimental
probabilities'' (frequencies):
$$P_{++}^{\rm{exp}}(\theta_1, \theta_2), \; P_{+-}^{\rm{exp}}(\theta_1, \theta_2),\;
P_{-+}^{\rm{exp}}(\theta_1, \theta_2), \;
P_{--}^{\rm{exp}}(\theta_1, \theta_2)$$ from the predictions of
quantum mechanics, namely,
$$P_{++}(\theta_1, \theta_2)=P_{--}(\theta_1, \theta_2)=
{1/2}\cos^2(\theta_1-\theta_2)$$ and $$P_{+-}(\theta_1,
\theta_2)=P_{-+}(\theta_1, \theta_2)={1/2}\sin^2(\theta_1-
\theta_2).$$ However, in some mysterious way those deviations
compensate each other and finally the correlation
$$
E^{\rm{exp}}(\theta_1, \theta_2)= P_{++}^{\rm{exp}}(\theta_1, \theta_2)-
P_{+-}^{\rm{exp}}(\theta_1, \theta_2)- P_{-+}^{\rm{exp}}(\theta_1,
\theta_2)+ P_{--}^{\rm{exp}}(\theta_1, \theta_2)
$$ is in the
complete agreement with the QM-prediction, namely, $$E(\theta_1,
\theta_2)= P_{++}(\theta_1, \theta_2)- P_{+-}(\theta_1, \theta_2)-
P_{-+}(\theta_1, \theta_2)+ P_{--}(\theta_1, \theta_2)= \cos
2(\theta_1- \theta_2).$$ Therefore such anomalies play no role in
the Bell's inequality framework. Nevertheless, other linear
combinations of experimental probabilities do not have such a
compensation property. There can be found non-negligible
deviations from predictions of quantum mechanics. Thus neither
classical nor quantum model can pass the whole family of
statistical tests given by all possible linear combinations of the
EPR-Bohm probabilities.

\medskip

Does it mean that both models are wrong?

\section{Eberhard-Bell Theorem}

In quantum information community  rather common opinion is that one could
completely exclude probability distributions from
derivation of Bell's inequality and proceed by operating with frequencies. One typically refers to the
result of works \cite{EB}--\cite{EB2} which we shall call the Eberhard-Bell theorem
(in fact, the first frequency derivation of Bell's inequality
was done by Stapp\cite{ST}, thus it may be better to speak about Bell-Stapp-Eberhard theorem).
 By this theorem Bell's inequality
can be obtain only under assumptions of realism -- the maps $\lambda \to a_\theta(\lambda) $ is well defined --
and locality -- the random variable $a_\theta(\lambda)$ does not depend on other variables which are
measured simultaneously with it. Thus (in opposite to the original Bell derivation) existence of
the probability measure $P$ serving for all polarization (or spin) projections is not assumed.

At the first sight it seems that our previous considerations  have no relation to the Eberhard-Bell theorem.
One might say: ``Yes, Bell proceeded wrongly, but his arguments are still true, because they were
justified by   Eberhard in the frequency framework.''

As was shown \cite{KHI}, the use of frequencies, instead of probabilities, does not improve Bell's considerations,
see also Hess and Philipp \cite{HPL2}.
The contextual structure of the EPR-Bohm experiment plays again the crucial role. If we go into details
of Eberhard's proof, we shall immediately see that he operated with statistical data obtained from three
different experimental contexts, $C_1,C_2, C_3,$ in such a way as it was obtained on  the basis of a single
context. He took results belonging to one experimental setup and add or substract them from results
belonging to another experimental setup.  These are not proper manipulations from the viewpoint of statistics.
One never performs algebraic mixing of data obtained for totally different sample.  Thus if one wants to proceed
in    Eberhard's framework, he should find some strong reasons that the situation in the EPR-Bohm experiment
differs crucially from the general situation in statistical experiments. I do not see such reasons. Moreover,
the EPR-Bohm experimental setup is very common from the general statistical viewpoint.

Moreover,   Eberhard's framework pointed to an additional source of nonexistence of a
single probability distribution, see De Baere \cite{Bae} and also \cite{DM}--\cite{DM2}.
Even if we ignore the contribution of measurement devices,
then the $\psi$-function still need not determine a single probability distribution. In
Eberhard's framework we should operate with results which are obtained in different runs. One could ask:
Is it possible to guarantee that different runs of experiment produce
 the same probability distribution
of hidden parameters? It seems that there are no reasons for such an assumption. We are not able to control
the source on the level of hidden variables. It may be that the $\psi$-function is just a symbolic
representation of the source, but it represents a huge ensemble of probability distributions of hidden
variables. If e.g. hidden variables are given by classical fields, see e.g. \cite{KHF1}--\cite{KHF3},
then a finite run of realizations (emissions of entangled photons) may be, but may be not
representative for the ensemble of hidden variables produced by the source.

\section{Comparing of the EPR and the EPR-Bohm experiments}

Typically the original EPR experiment \cite{E} for correlations of coordinates and momenta and the EPR-Bohm
experiment for spin (or polarization) projections are not sharply distinguished. People are almost sure
that it is the same story, but the experimental setup was modified to move from ``gedanken experiment''
to real physical experiment. However, it was not the case! We should sharply distinguish these two
experimental frameworks.

The crucial difference between the original EPR experiment and a new experiment
which was proposed by Bohm is that these experiments are based on quantum states
having essentially different properties. The original EPR state
$$
\Psi(x_1,x_2)=\int_{-\infty}^\infty \exp\left\{\frac{i}{\hbar}
(x_1-x_2+x_0) p \right\} d p,
$$
and the  singlet state
$$
\psi= \frac{1}{\sqrt{2}}(\vert + >\vert -> - \vert - >\vert + >)
$$
which is used in the EPR-Bohm experiment have in common
only one thing: they  describe correlated (or by using the modern terminology entangled)
systems. But, in contrast to the EPR-Bohm state, one can really (as EPR claimed) associate with the original EPR
state a single probability measure describing incompatible quantum observables (position and momentum).
The rigorous  prove in probabilistic terms was proposed by the author and Igor Volovich in \cite{KV}.
On the other hand, as we have seen for the singlet state one could not construct a probabilistic
model describing elements of reality corresponding to incompatible observables.

Thus the original EPR state is really exceptional from the general viewpoint of statistical analysis.
But the EPR-Bohm state behaves ``normally.'' In fact, there is no clear physical explanation
why statistical data for incompatible contexts can be based on a single Kolmogorov space in one case and not
in another. One possible explanation is that ``nice probabilistic features of the original EPR-experiment''
arise only due to the fact that it is ``gedanken experiment.''

\section{Legget's inequality and tests for nonlocal realistic
theories}

Here we  follow Legget \cite{Legget} and Gr\"oblacher et
al.\cite{Grob}, so details can be found in cited papers. The
source is assumed to distribute pairs of well polarized photons.
The size of the sub-ensemble in which photons have polarizations
$u$ and $v,$ respectively, is described by the density $F(u,v).$
Individual measurement outcomes are determined by a hidden
variable $\lambda$ (which may have a huge dimension; in fact, it
may belong to an infinite-dimensional space -- e.g. for a
classical field type hidden variables, see \cite{KHY}). For the
fixed polarizations $u$ and $v,$ the density of hidden variables
is given by $\rho_{u,v}(\lambda).$

The dichotomous $(\pm)$ measurement results are given by random
variables $A(a,b, u, v, \lambda)$ and $B(a,b, u, v, \lambda),$
where $a$ and $b$ are settings of polarization beam splitters of
Alice and Bob, respectively. The main Legget's trick (which was
repeated by Gr\"oblacher et al.) is that the average of  $A B$ is
calculated into the two steps:

\medskip

1). The average with respect to $\rho_{u,v}(\lambda).$

2). The average with respect to $F(u,v).$

\medskip

By using after the first step  some algebraic manipulations and
then averaging according to the second step, Legget obtained a
Bell-type inequality. Similar inequality was considered in
\cite{Grob} and tested experimentally.

The main problem of this derivation is that, instead of the
rigorous mathematical operation with conditional densities, we see
formal manipulation with densities $F(u,v)$ and
$\rho_{u,v}(\lambda).$ What is the real meaning of
$\rho_{u,v}(\lambda)$ in the rigorous mathematical framework? This
is nothing else than the {\it conditional density} $\rho(\lambda
\vert u, v).$  If one takes this fact into account, it would be
immediately clear that Legget's derivation suffers of the same
problem as the Bell's original one. It could be possible only
under the assumption of probabilistic compatibility of random
variables $A(a,b, u, v, \lambda), B(a,b, u, v, \lambda)$ for all
settings $a,b$ involved in considerations.

To simplify presentation, let us consider discrete  hidden
variable $\lambda$ and some discrete sampling of polarizations $u$
and $v$ (the latter is consistent with \cite{Grob}). Thus Legget's
considerations have the form. Set:
\begin{equation}
\label{ERT} \overline{AB} (u,v)\equiv \sum_{\lambda} A(a,b, u, v,
\lambda), B(a,b, u, v, \lambda) \; \rho_{u,v}(\lambda)
\end{equation}
and
\begin{equation}
\label{ERT1} \langle AB \rangle = \sum_{u,v} \overline{AB} (u,v)\;
F(u,v).
\end{equation}
Thus \begin{equation} \label{ERT2} \langle AB \rangle = \sum_{u,v,
\lambda} A(a,b, u, v, \lambda), B(a,b, u, v, \lambda) \;
\rho_{u,v}(\lambda)\; F(u,v).
\end{equation}
The crucial point of Legget's considerations is that he assumes
that the latter expression coincides with the classical
probabilistic  average $E(AB)$ of the product $AB.$ However, the
latter is valid only under assumption that there exists a
probability distribution $P(u,v, \lambda)$ such that
$$
F(u,v)= \sum_{\lambda} P(u,v, \lambda),
$$
is the marginal probability and
$$
\rho_{u,v}(\lambda)\equiv \rho(\lambda \vert u, v)= \frac{P(u,v,
\lambda)}{F(u,v)}
$$
is the conditional probability. Under such assumptions
\begin{equation}
\label{ERT3} \langle AB \rangle = \sum_{u,v, \lambda} A(a,b, u, v,
\lambda), B(a,b, u, v, \lambda) \; \rho_{u,v}(\lambda)\; F(u,v)
\end{equation}
$$
= E(AB)= \sum_{u,v, \lambda} A(a,b, u, v, \lambda), B(a,b, u, v,
\lambda) \; P(u,v, \lambda).
$$
Thus Legget's derivation is based on the (implicit) assumption:
existence of the probability distribution $P(u,v, \lambda).$
Moreover, to proceed further to his inequality Legget (as well as
Gr\"oblacher et al.\cite{Grob}) should assume that $P(u,v,
\lambda)$ does not depend on settings $a$ and $b.$ Thus they again
assume the probabilistic compatibility of random variables $A(a,b,
u, v, \lambda), B(a,b, u, v, \lambda)$ for a family of settings
$a$ and $b.$ We again do not see any physical or statistical
reason for such an assumption.

\section{Appendix: Proofs}

\subsection{Bell's inequality}

Let  ${\cal P}=(\Lambda, {\cal F}, P)$ be a Kolmogorov
probability space: $\Lambda$ is the set of parameters, ${\cal F}$ is a $\sigma$-algebra of its subsets
(used to define a probability measure), $P$ is a probability measure.
 For any pair of random variables $u(\lambda),
v(\lambda),$ their covariation is defined  by
$$
<u,v> = \rm{cov}(u,v)= \int_\Lambda u(\lambda)\;  v(\lambda) \; d {\bf
P}(\lambda).
$$
We reproduce the proof of Bell's inequality in the
measure-theoretic framework.

\medskip

{\bf Theorem.} (Bell inequality for covariations) {\it Let
$a, b, c= \pm 1$ be random variables on ${\cal P}.$  Then Bell's
inequality
\begin{equation}
\label{BBB} \vert <a,b > - < c,b >\vert \leq 1 - <a,c>
\end{equation}
holds.}

{\bf Proof.} Set $\Delta= <a,b > - < c,b >.$ By linearity of
Lebesgue integral we obtain
\begin{equation}
\label{B1} \Delta = \int_\Lambda a(\lambda) b(\lambda) d {\bf
P}(\lambda)- \int_\Lambda c(\lambda) b(\lambda) d P(\lambda)
\end{equation}
$$
=
\int_\Lambda [a(\lambda) - c(\lambda)]b(\lambda) d P(\lambda).
$$
As
\begin{equation}
\label{LLB}a(\lambda)^2= 1,
\end{equation}
we have:
\begin{equation}
\label{B2} \vert \Delta \vert = \vert \int_\Lambda [1 - a(\lambda)
c(\lambda)] a(\lambda) b(\lambda) d P(\lambda)\vert
\end{equation}
$$
\leq \int_\Lambda [1 - a(\lambda) c(\lambda)]  d P(\lambda).
$$

It is evident that ``hidden Bell's postulate'' on the existence of a single probability
measure $P$ serving for three different experimental contexts (probabilistic compatibility of three
random variables) plays the crucial role in derivation of Bell's inequality.

\subsection{Wigner inequality}

We recall the following simple mathematical result, see Wigner
\cite{Wig}:

\medskip

{\bf Theorem 1.2.} (Wigner inequality) {\it Let $a, b, c=\pm 1$ be
arbitrary random variables on a Kolmogorov space ${\cal P}.$ Then
the following inequality holds:}
\begin{equation}
\label{BB} P (a=+1, b=+1) + P(b=-1, c=+1)
\end{equation}
$$\geq {\bf
P}(a=+1, c=+1).
$$

{\bf Proof.} We have:
\begin{equation}
\label{W1}
\begin{array}{rl}
& P(  a(\lambda)=+1, b(\lambda)=+1)
\\ & \\
& =P(  a(\lambda)=+1, b(\lambda)=+1,
c(\lambda)=+1  )
\\ & \\
& + P(  a(\lambda)=+1, b(\lambda)=+1,
c(\lambda)=-1 ),
\\
\end{array}
\end{equation}
\begin{equation}
\label{W2}
\begin{array}{rl}
& P(  b(\lambda)=-1, c(\lambda)=+1)
\\ & \\
& =P(  a(\lambda)=+1, b(\lambda)=-1,
c(\lambda)=+1  )
\\ & \\
& + P(\lambda \in \Lambda : a(\lambda)= -1, b(\lambda)= -1,
c(\lambda)= +1),
\\
\end{array}
\end{equation}
and
\begin{equation}
\label{W3}
\begin{array}{rl}
& P(  a(\lambda)=+1, c(\lambda)=+1 )
\\ & \\
& =P(  a(\lambda)=+1, b(\lambda)=+1,
c(\lambda)=+1  )
\\ & \\
& + P(  a(\lambda)=+1, b(\lambda)=-1,
c(\lambda)=+1).
\\
\end{array}
\end{equation}
If we add together the equations (\ref{W1}) and (\ref{W2}) we obtain
\begin{equation}
\label{W4}
\begin{array}{rl} & P(   a(\lambda)=+1,
b(\lambda)=+1) + P(  b(\lambda)=-1,
c(\lambda)=+1)
\\ & \\
& =P (  a(\lambda)=+1, b(\lambda)=+1,
c(\lambda)=+1)
\\ & \\
& +P(  a(\lambda)=+1, b(\lambda)=+1,
c(\lambda)=-1)
\\ & \\
& +P(  a(\lambda)=+1, b(\lambda)=-1,
c(\lambda)=+1)
\\ & \\
& +P(  a(\lambda)=-1, b(\lambda)=-1,
c(\lambda)=+1).
\\
\end{array}
\end{equation}
But the first and the third terms on the right hand side of this
equation are just those which when added together make up the term
$P(  a(\lambda)=+1, c(\lambda)=+1)$ (Kolmogorov
probability is additive). It therefore follows that:
\begin{equation}
\label{W5}
\begin{array}{rl}
& P(  a(\lambda)=+1, b(\lambda)=+1) + {\bf
P}(  b(\lambda)=-1, c(\lambda)=+1)
\\ & \\
& =P(  a(\lambda)=+1, c(\lambda)=+1)
\\ & \\
&
 +P(  a(\lambda)=+1, b(\lambda)=+1, c(\lambda)=-1)
\\ & \\
& +P(  a(\lambda)=-1, b(\lambda)=-1,
c(\lambda)=+1)
\\
\end{array}
\end{equation}
By using non negativity of probability we obtain the inequality:
\begin{equation}
\label{W6}
\begin{array}{rl}
& P(  a(\lambda)=+1, b(\lambda)=+1) + {\bf
P}(  b(\lambda)=-1, c(\lambda)=+1)
\\ & \\
& \geq P(  a(\lambda)=+1, c(\lambda)=+1)
\\
\end{array}
\end{equation}

It is evident that ``hidden Bell's postulate'' on the existence of a single probability
measure $P$ serving for three different experimental contexts (probabilistic compatibility of three
random variables) plays the crucial role in derivation of Wigner's inequality.

\section{Conclusion}
In probability theory Bell's type inequalities were studied during last hundred years as constraints
for probabilistic compatibility of families of random variables -- possibility to realize them
on a single probability space. In opposite to quantum physics, such arguments as nonlocality and
``death of reality'' were not involved in considerations. In particular, nonexistence of a single probability space
does not imply that the realistic description (a map $\lambda \to a(\lambda))$
is impossible to construct. Bell's type inequalities were considered as signs (sufficient conditions)
of impossibility to perform simultaneous measurement {\it all random variables} from a family under consideration.
Such an interpretation can be used for statistical data obtained in the  EPR-Bohm experiment for entangled photons.

In any event, Bell's inequality could not be used to obtain the
definite conclusion on the relation between quantum and classical
models.

\section*{Acknowledgment}

I am  thankful for discussions  to L. Accardi, S.
Albeverio, M. Appleby, L. E. Ballentine, V. Belavkin, L. Hardy, I.
Bengtsson, M. Bozejko, C. Fuchs, S. Gudder, K. Gustafson, I.
Helland, A. Holevo, R. Hudson, S. Kozyrev, E. Loubenets, M. Manko
and V. Manko, D. Petz, R. Schack, A. Caticha,
D. Aerts, S. Aerts, B. Coecke, A. Grib, P. Lahti,
 A. Aspect, D. Mermin,
L. Accardi, G. Adenier, W. De Baere,  W. De Muynck, K. Hess and W. Philipp,
R. Gill, D. Greenberger, J-A. Larsson,  A. Peres, I. Pitowsky, M. Scully, I. Volovich.


\begin{thebibliography}{99}

\bibitem{GEN} M. Genovese, {\it Phys. Rep.} {\bf 413} (2005) 319.

\bibitem{B} J. S. Bell, \emph{Speakable and unspeakable in quantum
mechanics.} Cambridge: Cambridge Univ. Press, 1987.

\bibitem{DS} B. d'Espagnat, \emph{Veiled Reality. An anlysis of present-day
quantum mechanical concepts.} Addison-Wesley, 1995.

\bibitem{Shim} A. Shimony, \emph{Search for a naturalistic world view.}
Cambridge: Cambridge Univ. Press, 1993.

\bibitem{CL} J. F. Clauser , M.A. Horne, A. Shimony, R. A. Holt, \emph{Phys.
Rev. Letters,} vol. 49, pp. 1804-1806, 1969.

\bibitem{CL1} J. F. Clauser,  A. Shimony, \emph{Rep. Progr. Phys.,} vol. 41, pp. 1881-1901, 1978.

\bibitem{Wig} E. P. Wigner, \emph{Am J. Phys.}, vol. 38, pp. 1005-1015, 1970.


\bibitem{AS} A. Aspect,  J. Dalibard,  G. Roger, \emph{Phys. Rev. Lett.,} vol.
49, pp. 1804-1807, 1982.

\bibitem{AST} A. Aspect, Trois Tests Expérimentaux des Inégalités de Bell par
mesure de corrélation de polarisation de photons, PhD thesis No.
2674, Université de Paris-Sud, Centre D'Orsay (1983)

\bibitem{AS1} D. Home,  F. Selleri, \emph{Nuovo Cim.
Rivista,} vol. 14, pp. 2--176, 1991.

\bibitem{ST} H. P. Stapp, \emph{Phys. Rev.} ser. D, vol. 3, pp. 1303-1320,
1971.

\bibitem{EB} P. H. Eberhard, \emph{Il Nuovo Cimento} ser. B, vol. 38, pp. 75-80, 1977.

\bibitem{EB1} P. H. Eberhard, \emph{Il Nuovo Cimento} B, vol. 46, pp. 392-419, 1978.

\bibitem{EB2} P. H. Eberhard, \emph{Phys. Rev. Letters,} vol. 49, pp. 1474-1477, 1982.

\bibitem{P} A. Peres,  \emph{Am. J. of Physics,} vol. 46, pp. 745-750, 1978.

\bibitem{Weihs0} G. Weihs, T. Jennewein, C. Simon, H. Weinfurter, and A. Zeilinger,
\emph{Phys. Rev. Lett.,} vol. 81, pp. 5039-5042, 1998.

\bibitem{Weihs} G, Weihs,  ``A test of Bell's inequality with spacelike
separation,'' in \emph{Proc. Conf.
Foundations of Probability and Physics-4,}  Melville,
NY: American Institute of Physics, Ser. Conference Proceedings, vol. 889, pp. 250-262,  2007.

\bibitem{KHRP} A. Yu. Khrennikov (ed.), \emph{Proc. Conf. Foundations of
Probability and Physics,}  Singapore: WSP, Ser.  Quantum Probability and White Noise
Analysis, vol. 13, 2001.

\bibitem{KHRP1} A. Yu. Khrennikov (ed.), \emph{Proc. Conf. Quantum
Theory: Reconsideration of Foundations, } V\"axj\"o: V\"axj\"o Univ. Press,
Ser. Math. Modeling, vol. 2,  2002.

\bibitem{KHRP2} A. Yu. Khrennikov (ed.),  \emph{Proc. Conf.
Foundations of Probability and Physics-2,} V\"axj\"o: V\"axj\"o Univ. Press,
Ser. Math. Modeling, vol. 5,  2003.

\bibitem{KHRP3} A. Yu. Khrennikov (ed.), \emph{Proc. Conf.
Foundations of Probability and Physics-3,}
Melville, NY: American Institute of Physics, Ser. Conference Proceedings, vol. 750,   2005.

\bibitem{KHRP4} G. Adenier,   A. Yu. Khrennikov,  and  Th.M. Nieuwenhuizen (eds.),
\emph{ Proc. Conf. Quantum Theory: Reconsideration of Foundations-3,}
Melville, NY: American Institute of Physics, Ser. Conference Proceedings, vol. 810,   2006.

\bibitem{KHRP5} G. Adenier, C. Fuchs, and A. Yu. Khrennikov,  (eds), \emph{Proc. Conf.
 Foundations of Probability and Physics-3,}
Melville, NY: American Institute of Physics, Ser. Conference Proceedings, vol. 889,   2007.

\bibitem{FEY} R. P. Feynman and A. R. Hibs, {\it Quantum mechanics and path
integrals,} Wiley, New York, 1965.

\bibitem{DIR} P. A. M.  Dirac, {\it The Principles of Quantum Mechanics,} Oxford
Univ. Press, 1930.

\bibitem{KHI} A. Yu. Khrennikov, \emph{Interpretations of Probability.} Utrecht/Tokyo: VSP
Int. Sc. Publishers, 1999 (second edition, 2004).

\bibitem{KOL} A. N. Kolmogoroff,  \emph{Grundbegriffe der Wahrscheinlichkeitsrech.}
Berlin: Springer Verlag, 1933; reprinted : \emph{Foundations of the
Probability Theory.}  New York: Chelsea Publ. Comp., 1956.

\bibitem{G1}  B. V.  Gnedenko, {\em The theory of probability.} New-York: Chelsea Publ.
Com., 1962

\bibitem{ACC} L. Accardi,  ``The probabilistic roots of the quantum
mechanical paradoxes,'' in  \emph{Proc. Conf. The wave--particle dualism. A tribute to
Louis de Broglie on his 90th Birthday}, Dordrecht: Reidel Publ. Company,
pp. 47--55, 1970.

\bibitem{ACC1} L. Accardi, \emph{Urne e Camaleoni: Dialogo sulla realta,
le leggi del caso e la teoria quantistica.} Rome: Il Saggiatore,
1997.

\bibitem{ACC2} L. Accardi,  ``Could one now convince Einstein?'' In \emph{
Proc. Conf. Quantum Theory: Reconsideration of Foundations-3,}
Melville, NY: American Institute of Physics, Ser. Conference
Proceedings, vol. 810, pp. 3-18,  2006.


\bibitem{FINE} A. Fine,  \emph{Phys. Rev. Lett.,} vol. 48, pp. 291--295, 1982.

\bibitem{GAR1} C. Garola and L. Solombrino, \emph{Found. Phys.}
{\bf 26}, 1121 (1996).

\bibitem{GAR2} C. Garola and L. Solombrino, \emph{Found. Phys.}
{\bf 26}, 1329 (1996).

\bibitem{GAR3} C. Garola, \emph{Found. Phys.}
{\bf 32}, 1597 (1996).

\bibitem{HPL2} K. Hess and W. Philipp, `` Bell's theorem: critique of proofs
with and without inequalities,'' in \emph{Proc. Conf. Foundations of Probability and Physics-3,}
Melville, NY: American Institute of Physics, Ser. Conference Proceedings, vol. 750, pp. 150-155,  2005.

\bibitem{KH1} A. Yu. Khrennikov,  \emph{Il Nuovo Cimento,}  ser. B vol. 115,  pp. 179-184, 1999.

\bibitem{KH2}  A. Yu. Khrennikov,  \emph{J. of Math. Physics,} vol. 41,  pp. 1768-1777, 2000.

\bibitem{KH3}  A. Yu. Khrennikov,  \emph{J. of Math. Physics,} vol. 41,  pp. 5934-5944, 2000.

\bibitem{KHF} A. Yu. Khrennikov,  {\it Found. of Phys.} {\bf 32}, 1159-1174
(2002).

\bibitem{Kup1} M. Kupczynski, {\it Phys. Lett A} {\bf 116}, 417-422 (1986).

\bibitem{Kup2} M. Kupczynski, {\it Phys. Lett A} {\bf 121}, 51-56
(1987).

\bibitem{Kup3} M. Kupczynski, {\it Phys. Lett A} {\bf 121},
205-210 (1987).


\bibitem{PI}  I. Pitowsky,  \emph{Phys. Rev. Lett.,} vol. 48, pp. 1299-1302, 1982.

\bibitem{PI1}  I. Pitowsky, \emph{Phys. Rev.} ser. D, vol.  27,  pp. 2316-2326, 1983.

\bibitem{RAST} P. Rastal, \emph{Found. of  Physics,} vol.  13,  pp. 555-575, 1983.


\bibitem{Klyshko1} D. N. Klyshko {\it Phys. Lett. A}  {\bf 172},  399-403 (1993); ibid
{\bf 176}, 415-420 (1993); ibid {\bf 218}, 119-127 (1996); ibid
{\bf 247},  261-266 (1998).

\bibitem{Klyshko2} D. N. Klyshko, {\it Laser Physics} {\bf 6}, 1056-1076 (1996).

\bibitem{Klyshko3} D. N. Klyshko, {\it Annals of New York Academy of Science} {\bf 755},  13-27 (1995).

\bibitem{Klyshko4} D. N. Klyshko, {\it Uspehi Fizicheskih Nauk} {\it 168},
975-1015 (1998).

\bibitem{Andreev1}  V. A. Andreev, V. I. Man'ko, {\it JETP LETTERS} {\bf 72}, 93-96 (2000).

\bibitem{Andreev2}  V. A. Andreev, V. I. Man'ko, {\it Theor. Math. Phys.} {\bf 140}, 1135-1145 (2004).

\bibitem{Andreev3}  V. A. Andreev, V. I. Man'ko,  {\it Phys. Lett. A}  {\bf 281},  278-288 (2001)

\bibitem{Andreev4} V. A. Andreev, V. I. Man'ko, O. V. Man'ko, et al., {\it Theor. Math. Phys.} {\bf 146}, 140-151 (2006).

\bibitem{Andreev5} V. A. Andreev, {\it J. Russian Laser Research} {\bf  27},  327-331 (2006).

\bibitem{Manko1} V. I. Manko,  {\it J. of Russian Laser Research} {\bf 17}  579-584 (1996).

\bibitem{Manko2} O. V. Manko and V.I. Manko,  {\it J. Russian Laser Research} {\bf 25}, 477-492 (2004).

\bibitem{Manko3} V. I. Manko and E. V. Shchukin, {\it J. Russian Laser Research} {\bf 22}, 545-560
(2001).

\bibitem{Manko4} M. A. Manko, V. I. Manko, R. V. Mendes, {\it J. Russian Laser Research} {\bf 27}, 507-532 (2006).

\bibitem{Manko5} Yu. M. Belousov, V.I. Manko, {\it Density matrix:
representations and applications in statistical mechanics, v. 1,
2,} MFTI-Publ.,  Moscow, 2004

\bibitem{Vol1} I. V. Volovich,  ``Quantum cryptography in space and Bell's theorem'',
in \emph{Proc. Conf. Foundations of Probability and Physics,}
edited by A. Yu. Khrennikov, Ser. Quantum Probability and White
Noise Analysis {\bf 13}, WSP, Singapore, 2001, pp.364-372

\bibitem{Vol2} I. V. Volovich, ``Towards quantum information theory
in space and time'', \emph{Proc. Conf. Quantum Theory:
Reconsideration of Foundations,} edited by A. Yu. Khrennikov, Ser.
Math. Modeling {\bf 2}, V\"axj\"o Univ. Press, V\"axj\"o, 2002,
pp. 423-440.

\bibitem{KV} A. Yu.  Khrennikov and I. V. Volovich,
\emph{Soft Computing}    {\bf 10},   521 - 529 (2005).



\bibitem{IPT1} I. Pitowsky,  ``From George Boole to John Bell: The Origins of Bell's Inequalities,''
in \emph{Proc. Conf. Bell's Theorem, Quantum theory and
Conceptions of the Universe,} Dordrecht: Kluwer, pp. 37-49,  1989.

\bibitem{IPT2} I. Pitowsky, ``Range Theorems for Quantum Probability and Entanglement,''
in  \emph{Proc. Conf. Quantum Theory: Reconsideration of Foundations,}
Vaxjo: Vaxjo University Press, pp. 299-308, 2002.

\bibitem{SOZ} C. Garola and S. Sozzo, quant-ph/0703260 (2007).

\bibitem{VR} N. N. Vorob'ev, \emph{Theory of Probability and its
Applications},  vol.  7,  pp. 147-162, 1962.


\bibitem{Muc} W. Muckenheim,  \emph{Phys. Rep.,} vol.  133,  pp. 338--401, 1986.

\bibitem{KHP} A. Yu. Khrennikov, \emph{ Dokl. Akad. Nauk SSSR,} ser. Matem., vol.  322,  pp. 1075--1079,  1992.

\bibitem{KHP1} A. Yu. Khrennikov, \emph{J. Math. Phys.,} vol.  32,  pp. 932--937, 1991.

\bibitem{KHP2} A. Yu. Khrennikov, \emph{Phys. Lett. A,} vol.
200,  pp. 119--223, 1995.

\bibitem{KHP3} A. Yu. Khrennikov, \emph{Physica A,} vol.  215,  pp. 577--587, vol.  1995.

\bibitem{KHP4} A. Yu. Khrennikov, \emph{Int. J. Theor. Phys.,} vol.  34,  pp. 2423--2434, 1995.

\bibitem{KHP5} A. Yu. Khrennikov, \emph{J. Math. Phys.,} vol.  36,  pp. 6625--6632, 1995.

\bibitem{KHP6} A. Yu. Khrennikov, \emph{$p$-adic valued distributions in mathematical physics.} Dordrecht: Kluwer
Academic Publishers, 1994.

\bibitem{KHP7} A.Yu. Khrennikov, \emph{Non-Archimedean analysis: quantum paradoxes, dynamical systems and
biological models.} Dordrecht: Kluwer
Academic Publishers, 1997.

\bibitem{Bae} W. De Baere, \emph{Lett. Nuovo Cimento,} vol.  39,  pp. 234-238,
1984.

\bibitem{DM} W. De Muynck and W. De Baere W.,
\emph{Ann. Israel Phys. Soc.,} vol.  12,  pp. 1-22, 1996.

\bibitem{DM1} W. De Muynck, W. De Baere and H. Marten,
\emph{Found. of Physics,} vol.  24,  pp. 1589--1663, 1994.

\bibitem{DM2} W. De Muynck, J. T. Stekelenborg,  \emph{Annalen der Physik,} vol.  45,
 pp. 222-234, 1988.

\bibitem{KHF1} A. Yu. Khrennikov,  \emph{J. Phys. A: Math. Gen.} vol.
38,  pp. 9051-9073, 2005.

\bibitem{KHF2} A. Yu. Khrennikov, \emph{Found. Phys. Lett.,}  vol.  18,  pp. 637-650,
2006.

\bibitem{KHF3} A. Yu. Khrennikov, \emph{Physics Letters A,}  vol.  357,  pp. 171-176, 2006.

\bibitem{E}  A. Einstein, B. Podolsky, N. Rosen, \emph{Phys. Rev.,} vol.  47, pp.
777--780, 1935.

\bibitem{KV} A. Yu.  Khrennikov and I. V. Volovich,
\emph{Soft Computing,}  vol.  10,   pp. 521 - 529, 2005.

\bibitem{PEA} P. Pearle, \emph{Phys.
Rev.} ser. D, vol. 2, pp. 1418-1428, 1970.

 \bibitem{GIS} N. Gisin and B. Gisin, \emph{Phys.
Lett.}  ser. A., vol. 260,  pp. 323-333, 1999.

\bibitem{JAL}  Jan-\r{A}ke Larsson, \emph{Quantum Paradoxes, Probability
Theory, and Change of Ensemble.} Link\"oping, Sweden: Link\"oping
Univ. Press,  2000.

\bibitem{Gill} R. D. Gill, The chaotic chamelion.
quant-ph/0307217.

\bibitem{KHINTER} A. Yu. Khrennikov, EPR-Bohm experiment and interference
of probabilities. {\it Foundations of Physics Letters}, {\bf 17},
691-700 (2004).

\bibitem{ED1} E. Waks, K. Inoue, W. D. Oliver, E. Diamanti, Y.
Yamamoto, {\it IEEE J. Slected Topics in Quantum Electronics},
{\bf 9}, 1502-1511 (2003).

\bibitem{ED2} S. Nam, A. J. Miller and D. Rosenberg,
{\it Nuclear Instruments and Methods in Physics Research Section
A: Accelerators, Spectrometers, Detectors and Associated
Equipment}  {\bf 520},  523-526 (2004).

\bibitem{AKH} G. Adenier, A. Khrennikov, ``Anomalies in EPR-Bell experiments,'' in
\emph{Proc. Conf. Quantum theory:
reconsideration of foundations---3,}
Melville, NY: American Institute of Physics, Ser. Conference Proceedings, vol. 810, pp. 283--293,  2006.

\bibitem{AKH1} G. Adenier, A. Khrennikov,
\emph{J. Phys. B: Atomic, Molecular and Optical Physics,} vol. 40 (1), pp. 131-141,
2007.

\bibitem{Legget} A. J. Legget, {\it Found. Phis.} {\bf 33}, 1469-1493.

\bibitem{Grob} S. Gr\"oblacher, T. Paterik, R. Kaltenbaek, C.
Brukner, M. Zukowski, A. Aspelmeyer, and A. Zeilinger, {\it
Nature} {\bf 446}, 871 (2007).

\bibitem{L1} A. Lande, {\it Foundations of quantum theory,} Yale Univ.
Press, 1955.

\bibitem{L2} A. Lande, {\it New foundations of quantum mechanics,}
Cambridge Univ. Press, Cambridge, 1965.

\bibitem{L3} W. E. Lamb, {\it The interpretation of quantum mechanics,}
Rinton Press, Inc, Princeton, NJ, 2001.

\bibitem{S1} E. Santos, {\it Phys. Lett. A} {\bf 101},
379--382 (1984).

\bibitem{S2} E. Santos,  Bell's theorem and the experiments: Increasing empirical support to local realism, quant-ph/0410193v1 (2004).

\bibitem{KR1} A. F. Kracklauer,  quant-ph/0610173 (2006).

\bibitem{KR2}  A. F. Kracklauer, {\it Found. Phys.
Lett.} {\bf 19} (6), 625-629 (2006).

\bibitem{BOULD} M. A. Rowe, D. Kielpinski, V. Meyer, C. A. Sackett, W. M.
Itano, C. Monroe and D. J. Wineland , {\it Nature} {\bf 409},
791-794 (2001).

\bibitem{KHY} Khrennikov A. Yu.,   {\it J. Phys. A: Math. Gen.}
{\bf 38}, 9051-9073 (2005);   {\it Found. Phys. Letters} {\bf 18},
637-650 (2005).


\end{thebibliography}
\end{document}